\documentstyle[12pt]{article}
\begin{document}
\title{AN ALTERNATIVE APPROACH TO THE STATIC SPHERICALLY SYMMETRIC, VACUUM  GLOBAL  SOLUTION TO THE EINSTEIN EQUATIONS}
\author{Luis Herrera$^{1}$\thanks{e-mail: lherrera@usal.es},
 and  Louis Witten$^{2}$\thanks{e-mail: lwittenw@gmail.com},\\
\small{$^1$ Instituto Universitario de Fisica Fundamental y Matematicas}\\
\small{Universidad de Salamanca, Salamanca, Spain}\\
\small{$^2$Department of Physics,} \\
\small{University of Florida, Gainesville, FL. 32611, USA.}\\
\\
}
\maketitle

\vspace{-0.5cm}

\begin{abstract}
We propose an alternative description of the Schwarzschild black hole based  on the requirement that the solution be static not only outside  the  horizon but also inside it. As a consequence of this assumption, we are led to a change of signature implying  a complex transformation of an angle variable. There is a ``phase transition'' on the surface $R=2m$, producing a change in the symmetry as we cross this surface. Some consequences of this situation on the motion of test particles are investigated.

\end{abstract}

\maketitle

\section{The Schwarzschild solution}
The Schwarzschild solution of the Einstein gravitational field equations for empty space is the unique static, spherically symmetric, asymptotically
flat solution. These statements have a gauge independent meaning. 

Indeed, by ``static'' is meant that the metric tensor satisfies
the Killing equations for the timelike vector field $\chi _{(0)}$
\begin{equation}{\mathcal{L}}_{\chi_{(0)} } g_{\alpha  \beta } =0 , \label{1cmh}
\end{equation}where ${\mathcal{L}}$ denotes the Lie derivative with respect to the vector field $\mathbf{\chi }_{(\mathbf{0})} = \partial _{\mathbf{t}}$, which, besides being timelike is also hypersurface--orthogonal. 

Then, adapting our time--like coordinate to
the vector $\chi _{(0)}$, we must demand:
\begin{equation}\frac{ \partial g_{\mu  \nu }}{ \partial t} =g_{0 a} =0 , \label{w1}
\end{equation}in the above, Greek indices run from $0$ to $3$, whereas Latin indices run from $1$ to $3$. 

On the other hand, spherical symmetry implies that the solution admits three additional Killing vectors,
describing invariance under rotations, namely:
\begin{eqnarray}
{\bf \chi_{(1)}}&=&\partial_{\phi},\qquad  {\bf \chi_{(2)}}=-\cos \phi \partial_{\theta}+\cot\theta \sin\phi \partial_{\phi}\nonumber \\ {\bf \chi_{(3)}}&=&\sin \phi \partial_{\theta}+\cot\theta \cos\phi \partial_{\phi}.
\label{2cmh}
\end{eqnarray}

Then, as is well known, the line element can be written in polar coordinates in the form ($R >2 m$)
\begin{eqnarray}
ds^2&=&-\left(1-\frac{2m}{R}\right)dt^2+\frac{dR^2}{\left(1-\frac{2m}{R}\right)}+R^2d\Omega^2, \nonumber \\ d\Omega^2&=&d\theta^2+\sin^2 \theta d\phi^2,
\label{w2}
\end{eqnarray}
where the time coordinate $t$ is adapted to the timelike Killing vector (\ref{1cmh}). This is a static solution with the ($\theta, \phi$) curved 2-surface describing a space with a positive Gaussian curvature.

When $R <2 m ;$ the signature changes from (-, +, +, +) to (+, -, +, +). Although mathematically a solution, this does not describe
a physical solution as we wish it to be described. 

Since there is no algebraic singularity at the horizon ($R =2 m$), (the physically meaningful components of the Riemann tensor are regular), there are coordinate transformations that yield acceptable
solutions of the Einstein equations with an unchanged signature over the entire resultant space-time. These coordinate transformations yield acceptable
solutions of the Einstein equations over the entire manifold but they are not static. 

As stressed by Rosen \cite{rosen},
any transformation that maintains the static form of the Schwarzschild metric is unable to remove the singularity in the line element. Indeed,
various coordinate transformations have been found that allow the manifold to extend over the whole region $0 <R <\infty $, although all of them necessarily are non-static (for $R <2 m$) and require all incoming geodesics (test particles) to end at the central singularity (see for example \cite{eddington},
\cite{1}, \cite{fin},
\cite{krus}, \cite{is}, \cite{gau}).

The point is that for (\ref{w2}), $R =2 m$ is also a Killing horizon, implying that the timelike Killing vector $\mathbf{\chi }_{(\mathbf{0})}$ becomes null on that surface, and, still worse, becomes spacelike, inside it. 

From a physical point of view,
the existence of a static solution would be expected over the whole space time of a vacuum, as an equilibrium final state of a physical process is expected
to be static. Furthermore, we should recall that in obtaining the Schwarzschild solution, being static is not an additional assumption, but follows from
the sphericity and the vacuum condition. 

In this work we deviate from the classical approach which consists in extending the solution outside the horizon to the interior. We recognize that this is a legitimate possibility but the resultant space-time is necessarily not static. We insist on a static solution everywhere and this will require a radically different description of the significance of the horizon. The resultant static solution describes the space time as consisting of a complete four dimensional manifold on the exterior side and a second complete four dimensional solution in the interior. The two dimensional R-t sub manifold forms a single continuous topological manifold; the addition of a metric requires a change in signature at the horizon. The $\theta-\phi$ sub-manifolds have a spherical symmetry on the exterior and hyperbolic symmetry in the interior. The two meet only at a single curve, $R=2m$, $\theta=0$. Passage of a geodesic from one side of the $R-t$ manifold to the other is possible only on a point on this curve. The horizon has significance as a horizon only for static geodesic observers. A geodesic particle may start at $R=\infty$ and travel a distance to a position $R_{min}$ and remain there or return to $R=\infty$. $R_{min}$ may be interior or exterior to the horizon position $R=2m$, depending on the energy of the particle; for great enough energy, it will proceed all the way to $R=0$. The horizon has no significance for such a particle. Similarly a geodesic particle may travel from $R=0$ to $R_{max}$ and remain there or return to $R=0$ and $R_{max }$ will be related to the particles energy and not to $R=2m$. The particles may pass $R=2m$ as described only at $\theta=0$. In this paper we do not discuss the motions when $\theta$ is not $=0$.

\section{The solution inside the horizon}
Our main point is that one should certainly expect that static vacuum should be allowed inside the horizon as well as outside and one should
be able to describe a manifold that permits this. As we shall show in this note, this is in fact the case, although for doing that, we should permit for
a change in symmetry, from elliptic (spherical) outside the horizon to hyperbolic inside the horizon. This requires the transformation $d \Omega ^{2} \rightarrow  -d \Omega ^{2}$ which obviously calls for a detour through the complex plane. Doing so, we have a static solution everywhere, but the symmetry of the
$R =2 m$ surface is different at both sides of it. Furthermore, as we stated before, we have to consider two different manifolds at both sides of the horizon. Accordingly no match of both solutions at $R =2 m$ surface is needed.

Thus, for $R <2 m$ we have:
\begin{eqnarray}
ds^2&=&\left(\frac{2m}{R}-1\right)dt^2-\frac{dR^2}{\left(\frac{2m}{R}-1\right)}-R^2d\Omega^2, \nonumber \\ d\Omega^2&=&d\theta^2+\sinh^2 \theta d\phi^2.
\label{w3}
\end{eqnarray}

Just as when $R>2m$, this is also a static solution with the $(\theta  ,\phi )$ space describing a positive Gaussian curvature, and with the time coordinate $t$ adapted to the Killing vector (\ref{1cmh}). 

The above
metric obviously admits the time--like Killing vector $\mathbf{\chi }_{(\mathbf{0})} = \partial _{\mathbf{t}}$, and three additional Killing vectors which are:
\begin{eqnarray}
{\bf \chi_{(1)}}&=&\partial_{\phi},\qquad  {\bf \chi_{(2)}}=-\cos \phi \partial_{\theta}+\coth\theta \sin\phi \partial_{\phi},\nonumber \\ {\bf \chi_{(3)}}&=&\sin \phi \partial_{\theta}+\coth\theta \cos\phi \partial_{\phi}.
\label{2cmhb}
\end{eqnarray}

The above Killing vectors, of course, differ from the Killing vectors (\ref{2cmh}),
and define a different symmetry in the inner region. Such change of symmetry comes up through the transformation $\theta  \rightarrow i \theta $. 

The differences in the spatial $\theta  ,\phi $ subspaces for the external metric $(R >2 m)$ and the internal metric $(R <2 m)$ become clear when the 2--subspaces are embedded in a flat 3--space. 

Thus, for the external metric, we have:
\begin{equation}x =\sin  \theta  \cos  \phi  ;\quad \quad y =\sin  \theta  \sin  \phi  ;\quad \quad z =\cos  \theta  , \label{nw1}
\end{equation}with a Euclidean metric. 

The interior embedding is:
\begin{equation}x =\sinh  \theta  \cos  \phi  ;\quad \quad y =\sinh  \theta  \sin  \phi  ;\quad \quad z =\cosh  \theta  , \label{nw1}
\end{equation}with $ -1 =x^{2} +y^{2} -z^{2}$, $d s^{2} = -d x^{2} -d y^{2} +d z^{2}$. 

The exterior 2-surface is, as expected, a sphere. The interior 2-surface is a hyperboloid of two sheets. At
the horizon, they meet only at  $R =2 m$, $\theta  =0$. This intersection is not in the spacetime 4-d manifold because it is not contained in an open set within the manifold. 
Let
us now take a look at the motion of a test particle as a function of its own proper time. All particles have their proper time differences correlated, although
they are free to choose the starting proper time for their clocks. The paths in the two manifolds may be continuous across the horizon as the metric in
the $t -R$ plane has a continuous form. 

Obviously, motion from the exterior to the interior is possible only when
the two manifolds meet at $\theta  =0$. Thus we will assume radial motion. The entire plane is a single mathematical manifold but the change in signature would lead
one to expect that from the standpoint of physical observations, it represents two different manifolds, it may be called a single manifold with a ``phase
change'' at $R =2 m$. 

Of interest then is the path of a geodesic between $R =0$ and $R =\infty $. Whether we are interested in the path a particle takes in its proper time or in the observers time, we see that the velocity
will be given as a square. So wherever a particle went past a point going in one direction, it went past the same point going in the opposite direction
at some other time. It was, of course always traveling in the positive time direction, so it went past the same point while going in different directions
at two different times. 

At constant $\theta  =0$ the geodesics may be obtained, and because $ \partial _{t}$ is a Killing vector, there will be a constant of motion 

\begin{equation}\kappa  =(1 -\frac{2 m}{R}) \frac{ \partial t}{ \partial s} . \label{kw}
\end{equation}

From the metric we get the velocity
\begin{equation}
\left(\frac{dR}{ds}\right)^2=\kappa^2-1+\frac{2m}{R}
\label{18}
\end{equation}

Depending on $\kappa $ there are three possible solutions of (\ref{18}), stated parametrically:
\begin{enumerate}
\item $\kappa ^{2} -1 =0$.
\begin{eqnarray}
 R=(2m)^{1/3} \left(\frac{3}{2}\right)^{2/3} s^{2/3}.
\label{int4}
\end{eqnarray}

\item $\kappa ^{2} -1 >0$.
\begin{eqnarray}R =\frac{m}{\kappa ^{2} -1} (\cosh  \eta  -1) , &&\frac{m}{(\kappa ^{2} -1)^{3/2}} (\sinh  \eta  -\eta ) =s-s_0 \nonumber \\ s_0&=&\frac{\pi m}{(\vert \kappa ^{2} -1\vert )^{3/2}}. \label{int5}
\end{eqnarray} 

\item $\kappa ^{2} -1 <0$.
\begin{eqnarray}R& =&\frac{m}{\vert \kappa ^{2} -1\vert } (1 -\cos  \eta ) ,\frac{m}{(\vert \kappa ^{2} -1\vert )^{3/2}} (\sin  \eta  -\eta ) =s -s_0,\nonumber \\ s_0&=&-\frac{m}{(\vert \kappa ^{2} -1\vert )^{3/2}}(\sinh \tilde \eta-\tilde \eta), \quad \cos \tilde \eta=3, \label{int5e}
\end{eqnarray}\end{enumerate}
where for convenience the constants $s_0$ have been choosen such that particles coming from $R=0$ and $R=\infty$ meet  at $R=\frac{2m}{\vert \kappa ^{2} -1\vert}$, at the same time $s=0$.

The proper time, $s_{ ,}$  is taken to be $0$ when the particle comes to rest at $\frac{d R}{d s} =0$. As an aid in tracing the paths of geodesics, we see that all three cases have the same limiting behavior when the particle is at rest. All satisfy equation (\ref{18}) in this limit, independently of the value of $\kappa ^{2} -1$. 

We first discuss the motion of a particle with $\kappa ^{2} -1 <0$. We assume it enters at $R =0$. The behavior there is given by equation (\ref{int5e}), while the behavior of the velocity
as given by (\ref{18}), is $\frac{d R}{d s} \sim s^{ -1/3}$, so  the particle comes in to the origin at $\eta  =0$ and immediately curves around, to reemerge at once.  At $\eta  =\pi $ the particle is at its maximum distance, $R =\frac{2 m}{\vert \kappa ^{2} -1\vert }$, which may be larger or smaller than $2 m$. It comes to rest here, emerges on a cusp--like path and returns along the same path it entered.

In the above, we started out assuming the geodesic begins at $R =0$. Let us now consider the case when it begins at $R =\infty $. 

Then, for the radial geodesic at $\theta  =0$ we may write:
\begin{equation}1 =(2 m/R -1) (d t/d s)^{2} -1/(2 m/R -1) (d R/d s)^{2} , \label{n51}
\end{equation}denoting now by $q$ the constant of motion (the energy) associated to the time--like Killing vector and feeding it back into (\ref{n51}),
we get
\begin{equation}
\left(\frac{d R}{d s}\right)^{2} =q^{2} +1 -\frac{2 m}{R} . \label{n52}
\end{equation}\_

The geodesic comes in from infinity with a velocity $q^{2} +1$, reaches to a minimum zero velocity at $R =\frac{2 m}{q^{2} +1}$ and turns around immediately in a cusplike path and returns to infinity with the same speed as it left. 

The
end goal of starting both at $R =0$ and $R =\infty $, is to see what happens when two particles meet at the same point at the same time. Unfortunately, Einstein's theory does
not tell us if one geodesic bounces back to infinity and the other bounces back to $0$, or if each continues forward. In this latter case, the particle must gain energy or lose energy depending on which way it crosses $R =2m$.
They may also meet, interact,  and scatter and then return from whence they came. 

The particles will met when $s =0$ if   $\kappa ^{2} -1 =1 -q^{2}$.  If the vacuum level  of a zero energy particle in one manifold is  $ -1$, it may be $+1$ on the other side due to the change in signature, so $q^{2} = -\kappa $$\,^{2}$, and one may be the antiparticle of the other. If one is the antiparticle of the other, they may annihilate.

 Of course a full understanding of the picture described above requires  a quantum theory.

On the other hand let us look at the singularity $R =0$. It is the inverse curvature of a sphere and indicates that the curvature is infinite. However it stays at infinity for zero length of time. It is equivalent to a sphere reducing its radius very rapidly to $0$ and immediately, in zero time, expanding again. We would say that this singularity is physically  ignorable.

\section{Conclusions}
In \cite{rosen} the region $R <2 m$ was considered as non--physical and thereby excluded. According to Rosen, such an exclusion was justified by two reasons: on the one
hand because the coordinates $R$ and $t$ change from spacelike to timelike ($R$), and inversely $(t)$. On the other hand, he concluded  that particles reaching the surface
$R =2 m$ from the outside, are reflected back. 

In this work, we have considered the whole space--time. However, instead of sacrificing the staticity in the region inside the horizon (as in \cite{eddington, 1, fin, krus, is}),
we keep the time independence but change the spatial symmetry. 

For doing that we have to consider the whole space--time as consisting of  two distinct 4-manifolds, each of which corresponds to a static space--time. Each has the horizon as a boundary, each is a complete manifold. Complete means that a geodesic has a closed path or terminates at a boundary which is singular or infinite. One 4-manifold has the range $2m<R< \infty$ with the signature $(- + + +)$ and the second manifold has the range $0<R<2m$ with the signature $(+ - - - )$.  Proceeding in this way  we obtain a global static solution, which presents a ``phase transition'' on the
horizon, being endowed with different symmetry at both sides of it. Both manifolds   exclude the $R=2m$ surface. They meet at  $R=2m$, only when $\theta=0$. The acceptability of our setup is supported by the fact that  from purely physical arguments we  should expect a static solution to exist in the ``whole space--time'' (in the sense above).

Indeed, for $\theta=0$, as we have seen, the theory permits one geodesic to rebound to infinity and the other to bounce back
towards $0$. However each may continue in a straight path, one from $R =0$ to $R =\infty $ and the other in the opposite direction. One would gain two units of energy along the way and the other would lose two
units, $\kappa $ and $q$ refer to two different test particles. They may backscatter with different energies. The times of meeting may have been adjusted
so that the two particles meet at the same time, $s =0$. The zero energy of one vacuum is $ +1$ and, because of the change in signature, the zero energy of the other is $ -1$. Since at $s =0$, $\kappa ^{2} -1 =q^{2} +1$, the energy $\kappa $ is the negative of the energy $q$ and the two may be anti-particles.

If the particles were anti-particles, there would be another level of complexity and uncertainty
added to the mix. We have described the particle with the greater energy as falling from infinity and the one with the lesser as rising from $R =0$. It could just as well be the other way as far as the theory tells us. 

On the other hand, $R =0$, seems to be harmless. As an observable quantity, $R$ is a measure of the curvature of the surface it represents. A particle may, with a probability of measure 0 in a finite proper
time, reach a surface of zero curvature with infinite proper velocity and remain there for infinitesimal time. To verify that this statement is valid requires
a study of the geodesics including particles with non zero angular momentum, which we have not done here. 

The main lesson to extract from the comments above, is that there
is no way to understand the complete picture of the motion of test particles in the neighborhood of the horizon from Einstein's equations alone. A quantum theory is needed to understand it. The two curved surfaces of the
metric that meet at the horizon at $\theta  =0$ are continuous with continuous first derivatives but not second derivatives, so a particle cannot cross at the horizon except
at this one point. However, it is almost certain that a quantum theory would permit a particle to tunnel across.

It may look strange
that we have described a geodesic that comes in from infinity with the metric for $R <2 m$ . However this becomes intelligible if we recall that we are dealing only with $\theta  =0$ . The behavior is similar coming in from either direction. Either side could have been the starting point if $\kappa  =q =0.$ 

It should be pointed out that we have not considered the behavior of non radial geodesics. They
need major consideration as the non radial geodesics will be able to penetrate the horizon only by tunneling.

Also, it is worth noticing that we have not specified a source for the vacuum parameter $m$, which usually (as measured from the vacuum solution) is associated to a singular (point--like) source. However, an alternative origin for $m$ might be considered here. Indeed, we have already said  that we need quantum mechanics to understand the horizon. Then, if the vacuum has a ``mass'', the vacuum mass could come from  quantum mechanics. When students first learn about quantum mechanics and its zero point energy, they frequently ask why this does not show up in gravity. We speculate that our $m$ might be a quantum mechanics residue. 

Finally we would like to notice that an approach similar to the one presented here, could  be proposed for the global interpretation of the Kerr metric \cite{K}.

To summarize:
We have shown that there
is a static solution of Einstein's equations with the symmetry of a sphere or of a paraboloid and that the symmetry changes at the horizon. $R >2 m$ is a geodetically complete manifold in that each geodesic may be continued to a singularity or to a boundary. $R <2 m$ is also  a geodetically complete manifold. We recognize that our approach represents a radical departure from the standard description. The standard description insists on keeping the curved two surface involved ($\theta, \phi$ surface) unchanged but does not describe an everywhere vacuum, static situation. We have no difficulty with this description and furthermore are fully aware of the fact that it  had produced an impressive wealth  of research work ( just as a sample see \cite{bh1, bh7, bh3, bh5, bh6,  bh9, bh10, bh11, bh12} and references therein). However, we do believe  that a static vacuum at all times and places should be possible.

\section{acknowledgments}This work was partially supported by the Spanish Ministerio de Ciencia e Innovaci{\'o}n under Research Projects
No. FIS2015-65140-P (MINECO/FEDER).

\end{document}